\documentstyle[aps,preprint,tighten,psfig,twoside,floats]{revtex}
\begin{document}
\title{Scarring Effects on Tunneling in Chaotic Double-Well Potentials} 
\author{W. E. Bies\cite{bies-email}}
\address{Department of Physics, \\
  Harvard University, Cambridge, Massachusetts 02138}
\author{L. Kaplan}
\address{Institute for Nuclear Theory and Department of Physics, \\
  University of Washington, Seattle, Washington 98195}
\author{E. J. Heller}
\address{Department of Physics and Department of Chemistry and Chemical
  Biology, \\
  Harvard University, Cambridge, Massachusetts 02138}
\maketitle

\begin{abstract}%

The connection between scarring and tunneling in chaotic double-well
potentials is studied in detail through the distribution of level splittings.
The mean level splitting
is found to have oscillations as a function of energy, as
expected if scarring plays a role in determining the size of the
splittings, and the spacing between peaks is observed to be
periodic of period 
{$2\pi\hbar$} in action. Moreover, the size of the oscillations is
directly correlated with the strength of scarring. These results
are interpreted within the theoretical framework of Creagh and Whelan.
The semiclassical limit and finite-{$\hbar$} effects are discussed, and
connections are made with reaction rates and resonance widths in
metastable wells.

\end{abstract}
\newpage
\section{Introduction}
\label{intro}

Many chemical reactions must proceed through a potential barrier
before the final dissociation of the products of the reaction can
take place. Thus, the
reaction rate is governed by tunneling~\cite{chemical}.
Radioactive decay of nuclei also involves crossing a potential
barrier, dictated by a combination of short-ranged strong
binding forces and a longer-ranged Coulomb repulsion~\cite{nuclear}.
A typical experimental situation is the absorption of a slow neutron;
the resulting metastable nucleus decays predominantly by tunneling 
through a saddle point, and the distribution of resonance widths is well
described by random matrix theory \cite{PT}.
Another important example of quantum tunneling is the conductance of
mesoscopic devices~\cite{mesoscopic}.
In more than one dimension, when the classical dynamics 
can be chaotic, quantum chaos plays a role. For instance, in the
tunneling diode junction, in which electrons are driven by an
applied electric field but must tunnel through potential barriers
at either end of the device, the dynamics, in the presence of a
magnetic field, is chaotic and scarring of a short periodic orbit
is known to dominate the
conduction~\cite{tunneling-diode}. In this paper we study chaotic 
tunneling in a simple model
potential, and establish the connection to scar theory, both
qualitatively and quantitatively.
We prefer to study tunneling by calculating splittings in a double-well
potential rather than resonance widths in a metastable well,
because the former is an easier computational
task, as we will see later in Sec.~\ref{method}.
The relation between splittings in a double-well potential
and resonance widths in a metastable well will be
discussed immediately
below, where we will observe that the two quantities are related
by known overall normalization factors.

The double-well potential we will work with,
in two dimensions, has the very simple form
\begin{equation}
V(x,y)=x^4-x_0^2x^2+ay^2+\lambda x^2y^2+
\sum_i b_i e^{-((x-x_i)^2+(y-y_i)^2)/\sigma_i^2} \,.
\label{potential}
\end{equation}
The parameters {$x_0$}, {$a$}, {$\lambda$}, {$b_i$}, {$x_i$}, {$y_i$}
and {$\sigma_i$} will be specified below. 
For $x_0^2 > 0$, a barrier along the $y$-axis separates the potential 
at low energies ($E<0$) into two wells. The {$\lambda x^2y^2$} term ensures
that the potential is not separable, while the Gaussian perturbations, the
parameters of which may be changed at will, allow us to generate an ensemble
of statistically independent eigenstates near any given energy.
If the positions and heights of
the Gaussians are chosen suitably, the potential has reflection symmetries
in the {$x$} and {$y$} directions. The symmetry under reflection in {$y$}
means that there will always be a short periodic orbit on the {$x$}-axis,
in each of the two wells.

The principal result of this paper is that the size of the level splittings
in the two-dimensional double well, the classical dynamics of which 
is chaotic, is directly correlated with the scarring of eigenfunctions
along the {$x$}-axis, which we believe to be the primary channel for
tunneling. The evidence for this is that (1) the distribution of splittings,
once the average exponential trend in energy is scaled out, displays
oscillations as a function of energy of the sort expected by scar
theory. In particular, the action distance between successive peaks
is precisely {$h$}. However, the dependence of the mean splitting
on action is not well reproduced quantitatively by linear scar theory, because
non-linear effects appear to be important.
(2) The rescaled splittings are correlated with the
overlap of the eigenstate with a Gaussian test state lying on the
periodic orbit on the {$x$}-axis, a measure of scarring. 
(3) The correlation
of the splittings with the overlaps is further supported by the fact that
the distribution of overlaps has the same energy-dependent oscillations
as the distribution of splittings.

According to one-dimensional WKB theory, the splitting {$\Delta E
= E_{\rm anti-symm} - E_{\rm symm}$} in a symmetric double well
is given in the semiclassical limit by
\begin{equation}
\Delta E = \left( {{\hbar \omega} \over \pi} \right) e^{-S/\hbar} \,,
\label{dewkb}
\end{equation}
while the resonance width {$\Gamma$} for the state at the same energy
in a metastable well is
\begin{equation}
\Gamma = \left( {{\hbar \omega} \over {4\pi}} \right) e^{-2S/\hbar} \,.
\label{gwkb}
\end{equation}
Here {$\omega$} is the frequency of the classical periodic motion
at energy {$E_{\rm symm}$} and the imaginary action for going under the barrier
is 
\begin{equation}
S=\int_{-x_{\rm tp}}^{x_{\rm tp}} \, dx \,
\sqrt{V(x)-E} \,,
\end{equation}
{$x_{\rm tp}>0$} being the position of the 
classical turning point at energy {$E_{\rm symm}$}~\cite{wkb}. In order for the
semiclassical theory to apply, {$E_{\rm symm}$} must be sufficiently
far below the barrier that {$S/\hbar \gg 1$}. Note the factor of two
in the exponent in Eq.~(\ref{gwkb}).

These one-dimensional formulae may be generalized to the two-dimensional
(or higher-dimensional)
potential well as follows. We expect tunneling to
be dominated by paths that cross the barrier close to
the {$x$}-axis, which has the smallest action integral $S$.
The splitting {$\Delta E$} will then be proportional to the exponential
factor {$e^{-S/\hbar}$}. 

The correct generalization of the frequency of attempts
to cross the barrier, $\omega$, to two or more dimensions, 
is the frequency with which
one returns to a Planck-sized cell in phase space that lies on the horizontal
periodic orbit. This horizontal periodic orbit is the real continuation of
the least-action path across the barrier. The 
time for returning to
such a cell (or to any other cell in an ergodic well)
is the Heisenberg time {$T_H = h/\Delta(E)$}, where {$\Delta(E)$} is
the mean level spacing near energy $E$
(i.e., the spacing between doublets in the double-well system).
Then the frequency of attempts
to cross the barrier is just proportional to the mean level spacing
{$\Delta(E)$}.
Thus, we expect on general grounds
that the splitting {$\Delta E$} is given in order of magnitude by 
{$\Delta E \sim \Delta(E) e^{-S(E)/\hbar}$},
which gives us the trend of the splittings as a function of
energy. This expression for the mean splitting will be confirmed by the 
exact semiclassical theory to be discussed below. 

For any given state, we should expect that its splitting
will be large or small compared with the mean value
at that energy according to
whether its amplitude is large or small on the horizontal periodic orbit
which leads to optimal  tunneling. For simplicity we can study the wave
function amplitude near
the turning point of the horizontal periodic orbit.
The value of the wave function at the turning point in the two-dimensional
chaotic system is (ignoring scar-related effects which are the main
focus of this paper) given approximately
by a Gaussian-distributed random variable, as random matrix theory would
predict. Thus, {$|\psi(x_{\rm tp})|^2$} has, according to random matrix
theory, a Porter--Thomas distribution
for all energies far enough below the barrier (near zero energy the
horizontal periodic orbit becomes stable and the distribution of splittings
rolls over to one having many more large and small splittings, corresponding
to wave functions that live near or avoid this stable orbit).
However, what is relevant to tunneling in $d$ dimensions is not 
just the value of the wave function exactly at the 
turning point but rather its behavior in a whole $\hbar^{d-1}-$sized
region surrounding the periodic orbit.
We shall see below that the right quantity to consider
is the inner product of
the wave function with a Gaussian centered on the periodic orbit
(at the turning point or at some other location). This
has as well a Porter--Thomas distribution, within the random
matrix theory approximation.
Scar-related effects and finite-{$\hbar$} effects on the distribution of
splittings will be discussed below. 

The viewpoint summarized above is in agreement with the theoretical work of
Creagh and Whelan \cite{CW}. First, they find that the mean
splitting {$\langle \Delta E \rangle$} at a given energy
is given by the product of
an exponential
factor {$e^{-S/\hbar}$}, a factor proportional to the mean spacing
between doublets, as described above, and a
third factor that carries information about the monodromy matrix of the
(imaginary time)
tunneling orbit. 
Then they show that, for chaotic and
symmetric double wells, the splitting for a particular eigenvalue, relative
to the mean splitting,
may be written in the
semiclassical limit as a matrix element of the wave function {$\psi$} 
near the real continuation ${\cal R}$ of the complex trajectory that passes
through the barrier with minimum (imaginary) action. This imaginary-time
trajectory and its real-time continuation
may be thought of as the optimal route for tunneling.
The matrix element {$\Delta E \sim \langle \psi | {\cal T} | \psi \rangle$}
involves integration over a Poincar\'e surface
transverse to the real continuation ${\cal R}$. The kernel {${\cal T}$} is 
a semiclassical Green's function which, in the approximation that
the dominant contribution to the tunneling matrix element comes from
the neighborhood of ${\cal R}$,
becomes a Gaussian centered
on the intersection of ${\cal R}$ with the Poincar\'e section.
The width of the Gaussian is of {$O(\hbar^{1/2})$} in both directions
(e.g. $y$ and $p_y$) tangent to the surface of section. [The results may
of course be easily generalized to dimensions $d>2$, where the resulting
Gaussian has width of $O(\hbar^{1/2})$ in all $2d-2$ directions along the
surface of section.] 

In the case that the real continuation ${\cal R}$ happens to lie on a short
periodic orbit, which will always be true when a
reflection symmetry across the $x$-axis
is present, this matrix element may be regarded
as a measure of scarring on the periodic orbit. The Creagh--Whelan theory
predicts, therefore, that strong
scarring should be correlated with large splittings, confirming
the intuitive expectation that high tunneling rates should occur
for those wave functions that have large amplitude along the path
with optimal tunneling. Neither in Ref.~\cite{CW} nor in
Ref.~\cite{CW2}, however, do the authors demonstrate 
the connection between scarring and splittings on a state-by-state
basis. In the latter work, they confirm their formula for the
tunneling matrix element by deriving from it an analytical prediction
for the statistical distribution of splittings. This prediction is in agreement
with numerical calculations for potentials in which the real continuation
of the optimal tunneling orbit is {\it not} a periodic orbit; when it is,
the random-matrix assumption in their derivation breaks down due to
scarring on this periodic orbit. Thus, the present paper, while confirming
the predictions of Creagh and Whelan, goes beyond their results by
establishing conclusively the link between 
scarring and tunneling and by showing, with better statistics, that
the distribution of scaled splittings indeed
becomes approximately Porter--Thomas,
but only after scarring effects have been taken into account.

\section{Method}
\label{method}

The wave functions and splittings were calculated numerically
using the discrete variable
representation~\cite{dvr}.
The matrix elements of the position operators {$X$} and {$Y$} and of the 
kinetic energy operators {$K_x$} and {$K_y$} were first 
evaluated analytically using standard 
identities, in a basis of up
to the first 300 Gauss-Hermite functions in each dimension.
Then, in order to take advantage of the two reflection symmetries
in {$x$} and {$y$}, the two operators {$X^2$} and {$Y^2$} were diagonalized.
The reason for using $X^2$ and $Y^2$ instead of the
usual choice of
{$X$} and {$Y$} is that {$X^2$}, {$K_x$}, {$Y^2$} and {$K_y$} are all
block-diagonal, connecting only basis elements within one of the four
symmetry classes (even--even, even--odd, odd--even and odd--odd).
Since we are interested only in even--even potentials of
the form {$V(X,Y)=\sum_i f_i(X^2)g_i(Y^2)$} we can just as well compute
{$V$} at the eigenvalues of {$X^2$} and {$Y^2$} as at those of
{$X$} and {$Y$}, but using the basis obtained by diagonalizing
{$X^2$} and {$Y^2$} ensures that
the final Hamiltonian {$H=K_x+K_y+V$} 
will be itself also block diagonal. The
four symmetry classes may therefore be analyzed separately. In the
basis chosen the potential {$V$} is of course diagonal, while, in
two or more dimensions, the kinetic energy matrix will be sparse. 
More precisely, if {$N$} is
the dimension of one of the blocks in the Hamiltonian matrix, the
total number of non-zero matrix entries scales as $N^{1+1/d}$, or
{$N^{3/2}$}
in the two-dimensional system. Since we
require large values of $N$ in order to observe semiclassical
behavior, a sparse matrix routine is the method
of choice for diagonalizing {$H$}. The accuracy of the computed eigenvalues
was tested for convergence under increase of {$N$}, and for the parameters
given below we found convergence
to {$\pm 10^{-12}$} for {$N \approx 3500$}, corresponding to
about 200 Gauss-Hermite
functions in the {$x$} direction and about 100 in the {$y$} direction.

The amount of phase space covered by the region {$E<0$}, and thus the
number of states under the barrier and the 
computation time, increases very rapidly with {$x_0$}. If all other parameters
in the Hamiltonian are kept fixed, the number of states grows as $x_0^6$, and
so the largest value
of {$x_0$} we can easily attain is about {$x_0=6$}, for {$a=1$} and
{$\lambda=10$}. At these parameter values,
each well has a depth of {$x_0^4/4=324$} and about
100 bound states, where $\hbar$ is taken to be unity here and in the following.
The typical level spacing near the top of the
well is {$\sim 1$}, and the splittings range from {$<1$} near
the top of the well to {$<10^{-6}$} near {$E=-50$}; below this
energy many of the splittings become too small ({$<10^{-12}$})
to be resolved numerically. Therefore we take {$E=-50$} to be the
lower cutoff for the energies to be analyzed in Sec.~\ref{results}.

\section{Results}
\label{results}

\begin{figure}

\centerline{
\psfig{file=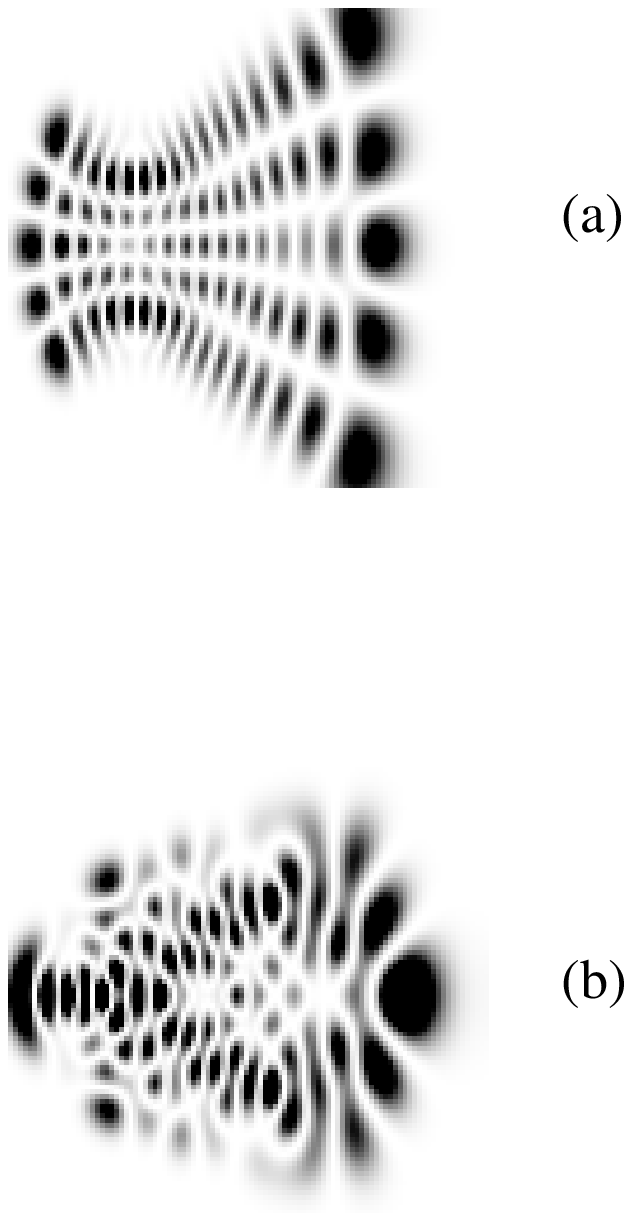,width=3in,bbllx=210pt,bblly=215pt,bburx=400pt,bbury=580pt,clip=}}

\vskip 0.2in

\caption{Typical eigenfunctions for the double-well potential in the (a)
near-integrable case without Gaussian perturbations, {$E=-11.055$}, and
(b) chaotic case with Gaussian perturbations, {$E=-12.063$}. Only one side
of the well is shown in each case, with the {$x$}-axis running horizontally
from $-6$ to 0 and the {$y$}-axis vertically from $-2$ to 2. The barrier is
located on the right side at {$x=0$}.}

\label{eigenfctns}
\end{figure}

The potential given by Eq.~(\ref{potential}), apart from the Gaussian
perturbation, is mostly integrable for all energies except those near
the top of the barrier. When the Gaussian perturbations are introduced,
the classical mechanics becomes more chaotic, but if
these perturbations
are too small it is still possible that
they would not be seen by the quantum
mechanics, which would remain effectively integrable. Thus, in order to
render the quantum mechanics at energies corresponding to the bound
states chaotic as well, it is necessary to introduce a Gaussian 
perturbation that is at least as large as the wavelength in question, and whose
height is comparable in magnitude to the depth of the potential well.
The simplest choice is to place large Gaussian perturbations above
the minima of the potential well at {$(\pm x_0/\sqrt{2},0)$}, which
will be seen by every bound trajectory as it crosses the center of the well
and which thus effectively make the dynamics chaotic at energies
down to the lowest considered ({$E=-50$}). This was checked classically
by examining the Poincar\'e surfaces of section and may also be seen to be
true quantum-mechanically in Fig.~\ref{eigenfctns}, where
typical eigenfunctions are shown.
Here, we have chosen the parameters of the double-well potential to be
{$x_0=6$}, {$a=1$} and {$\lambda=10$}, and for the central
Gaussian perturbation we use a height
{$b=150$} (to be compared with a well depth of {$V(\pm x_0/\sqrt{2},0)=-324$})
and a width {$\sigma=0.5$}.

\begin{figure}

\centerline{
\psfig{file=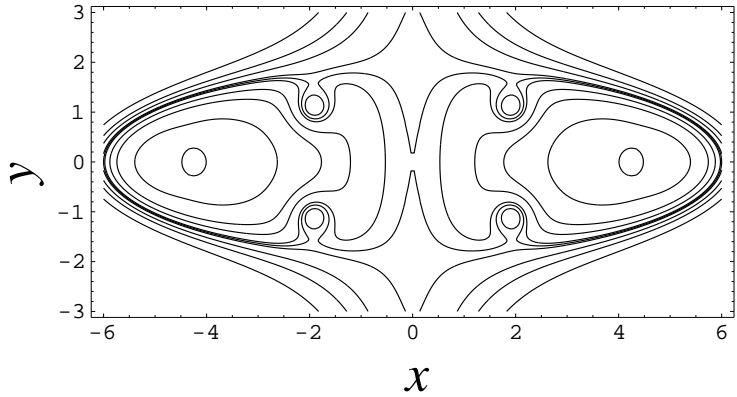,width=4in,bbllx=150pt,bblly=310pt,bburx=465pt,bbury=485pt,clip=}}

\vskip 0.2in

\caption{Contour plot of the potential for a typical member of the ensemble.
The perturbation at {$x=\pm 2, y=\pm 1$} leads to a symmetrical distortion
of the contours, which would be more rounded in the absence of a perturbation.
The contours range from {$V=-300$} near the bottom of the potential to
{$V=+200$} above the top of the barrier.}

\label{V}
\end{figure}

We generate an ensemble of 625 systems by placing four further
Gaussians (and their reflections in {$x$} and {$y$}) at 
{$x=\pm 2, \pm 3, \pm 4,  \pm 5$} and {$y=\pm 1$}, 
with heights {$b_i=20n_i$}, {$n_1,\ldots,n_4=1,\ldots,5$}, and with 
equal widths
$\sigma_i=0.5$ as above.
A contour plot of the potential for a typical member of the ensemble is given 
in Fig.~\ref{V}; note the symmetrical distortion of the contours due to the 
perturbation.

\begin{figure}

\centerline{
\psfig{file=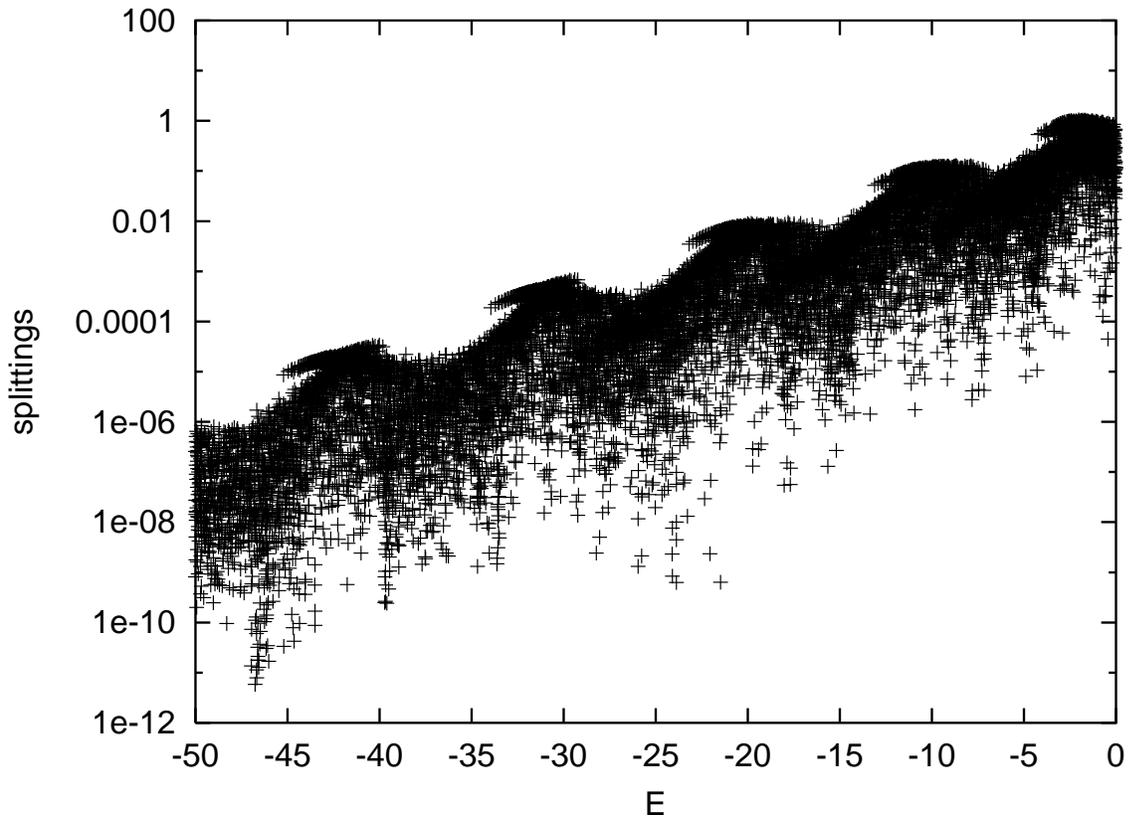,width=6in,bbllx=50pt,bblly=50pt,bburx=540pt,bbury=405pt,clip=}}

\vskip 0.2in

\caption{Level splitting versus energy {$E$} for the 15195 eigenstates 
between {$E=0$} and {$E=-50$} in the ensemble of 625 double-well potentials
described in the text.}
  
\label{se}
\end{figure}

\begin{figure}

\centerline{
\psfig{file=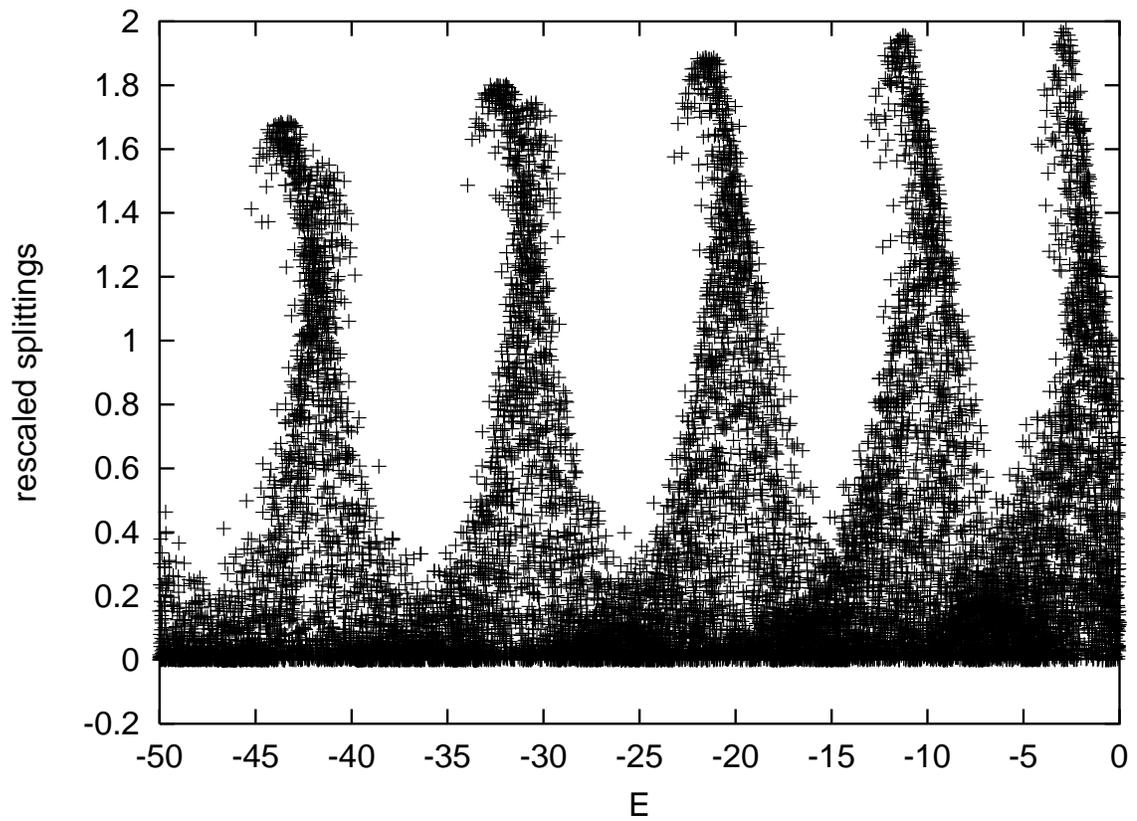,width=6in,bbllx=50pt,bblly=50pt,bburx=540pt,bbury=405pt,clip=}}

\vskip 0.2in

\caption{Rescaled level splitting versus energy {$E$} as
in Fig.~\ref{se}.} 
  
\label{serescaled}
\end{figure}

\begin{figure}

\centerline{
\psfig{file=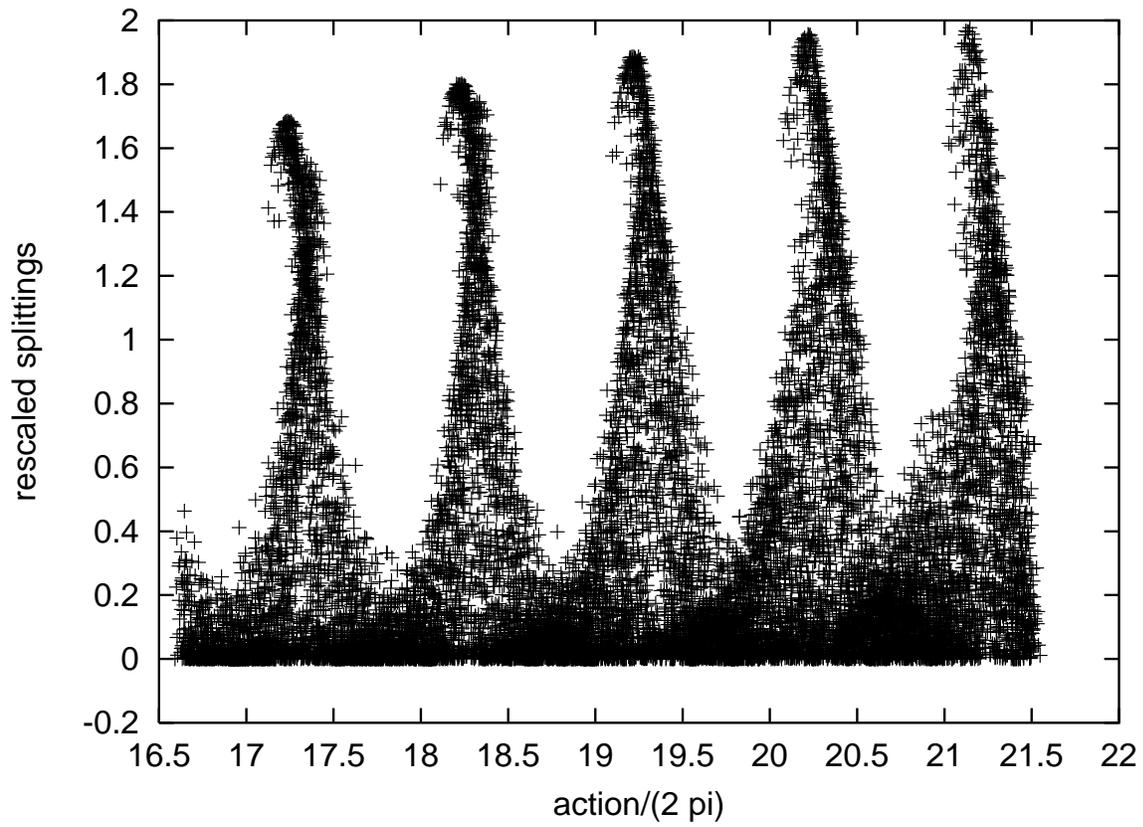,width=6in,bbllx=50pt,bblly=50pt,bburx=540pt,bbury=405pt,clip=}}

\vskip 0.2in

\caption{Rescaled level splitting versus action{$/2\pi$} with eigenstates as
in Fig.~\ref{se}.} 
  
\label{sa}
\end{figure}

We proceed to analyze statistically the splittings between states in the
even--even and even--odd sectors.
The results, for the parameters described, are given in Figs.~\ref{se}
and \ref{serescaled}. As expected, the size of the splittings
falls off exponentially with decreasing energy in
Fig.~\ref{se}, as the barrier becomes wider
and tunneling is suppressed. The trend is approximately linear on
a semi-log plot, over six decades as {$E$} varies from 0 to $-50$. 
In Fig.~\ref{serescaled} we rescale the splittings as a function of 
energy by {$s \rightarrow s/e^{-S(E)}$}.
We find very pronounced oscillations
in the distribution of splittings as a function of energy. As discussed
in Sec.~\ref{intro}, we expect theoretically that the rescaled splitting should
be proportional to the overlap in a Poincar\'e section
of the eigenfunction with a Gaussian on the horizontal periodic orbit. 
If the Gaussian may be assumed to have area exactly
{$h$} in the Poincar\'e section, 
such overlaps are described by scar theory \cite{KH,AF,LK}. We expect the 
results to be qualitatively
the same even if the Gaussian in the Poincar\'e section given by the theory
of Creagh and Whelan is not a minimum-uncertainty state, as long as it is
not too large compared to {$h$} (in the latter limit, scar effects must go
to zero in accord with the Schnirelman ergodicity theorem \cite{SCdVZ}). 
In fact, in the data presented below the area of the Creagh--Whelan
Gaussian ranged from {$1.5h$} to {$4h$}.
The prediction of scar theory is that, at a given
energy, the distribution of splittings should be Porter--Thomas, and that
the mean wave function intensity and therefore the mean splitting should 
oscillate as energy is varied by an amount that
depends on the Lyapunov exponent and the monodromy matrix
of the unstable periodic orbit.
An important confirmation of the scarring picture is obtained when we plot,
in Fig.~\ref{sa}, the rescaled splittings versus the action (divided by
{$2\pi$}) of the horizontal periodic orbit at the energy eigenvalue.
We find that the oscillations are periodic in action with period {$2\pi$},
which indicates that the EBK quantization condition for scarring holds.
The EBK quantization condition for the action reads {$A=
2\pi(n+1/2+n_c/4)$} where $n$ is an integer and {$n_c$} 
is the number of conjugate points in one period of the orbit
($n_c=3$ in the case of the  horizontal orbit in our system).

\begin{figure}

\centerline{
\psfig{file=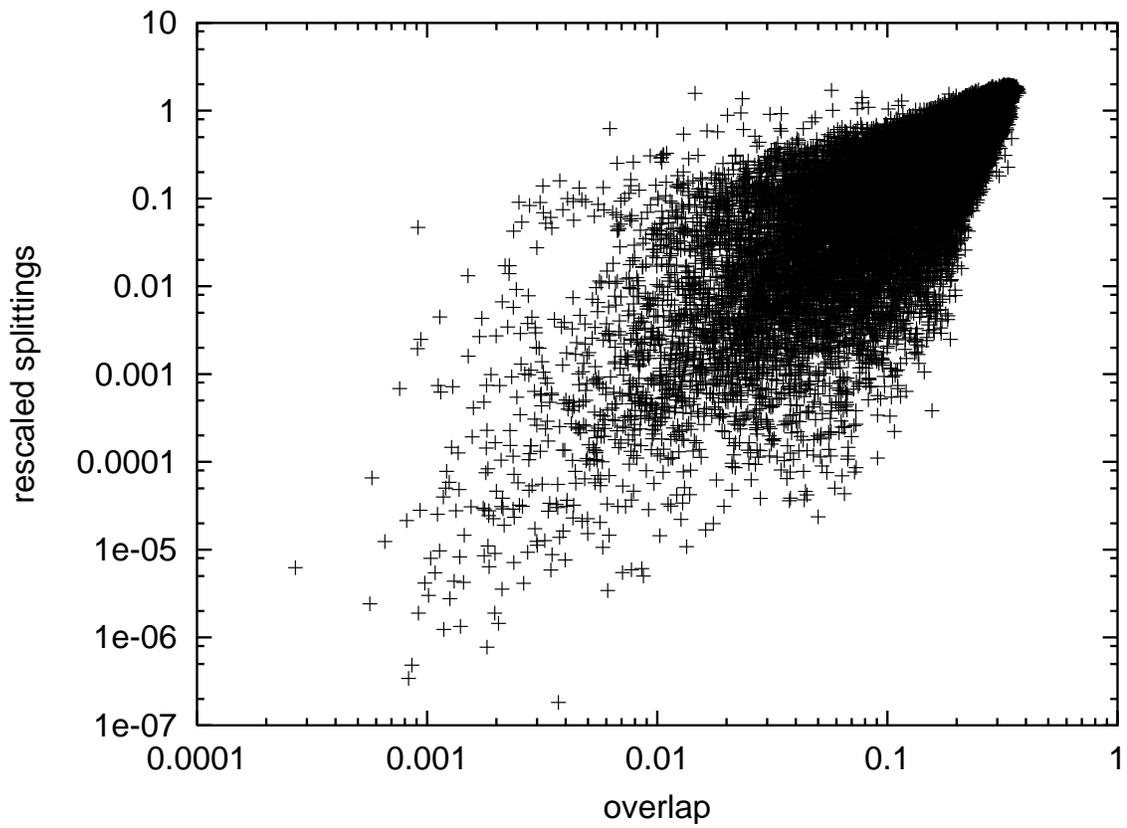,width=6in,bbllx=50pt,bblly=50pt,bburx=540pt,bbury=405pt,clip=}}

\vskip 0.2in

\caption{Rescaled level splitting versus the overlap of the eigenstate with
a Gaussian on the horizontal periodic orbit, with eigenstates as
in Fig.~\ref{se}.} 
  
\label{so}
\end{figure}

\begin{figure}

\centerline{
\psfig{file=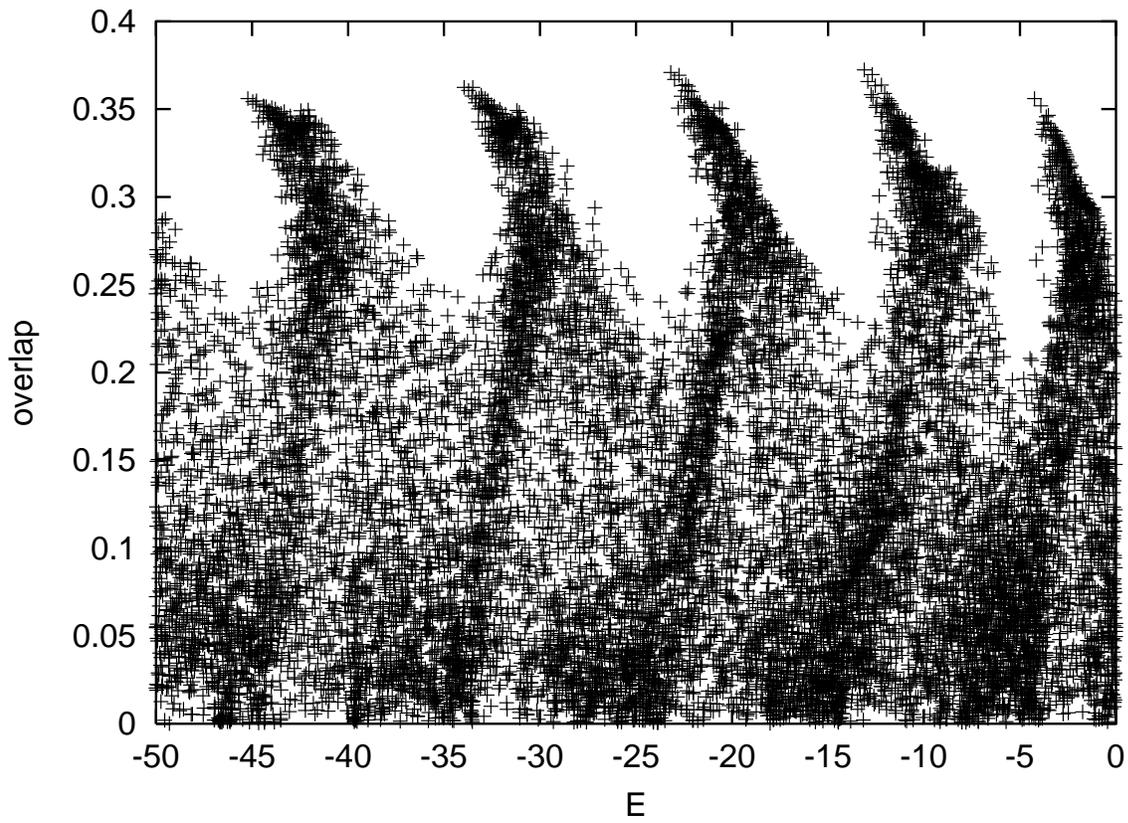,width=6in,bbllx=50pt,bblly=50pt,bburx=540pt,bbury=405pt,clip=}}

\vskip 0.2in

\caption{Overlap versus energy {$E$}, with eigenstates as
in Fig.~\ref{se}.} 
  
\label{oe}
\end{figure}

A direct correlation between splittings and scarring is found by
plotting, in Fig.~\ref{so}, the rescaled splitting of each eigenvalue versus
the overlap of the corresponding
eigenfunction with a Gaussian test state lying on
the horizontal periodic orbit, a measure of the degree with which this
eigenfunction is scarred. The two quantities are correlated, with a slope of 2
on a log-log scale.
The correlation coefficient of the logarithms
is 0.78; it may be that the degree of
correlation would be improved if the Gaussian were chosen to be properly
aligned with respect to the monodromy matrix of the optimal tunneling path.
The observed correlation
nevertheless confirms that there is a direct connection, on a 
state-by-state basis, between scarring and tunneling, as predicted by the 
theory of Creagh and Whelan \cite{CW}. 
As a check on our results, we show in Fig.~\ref{oe}
that the overlaps display the same energy-dependent oscillations as do the
splittings, as they must if the phenomenon of scarring underlies
the behavior of both.

\begin{figure}

\centerline{
\psfig{file=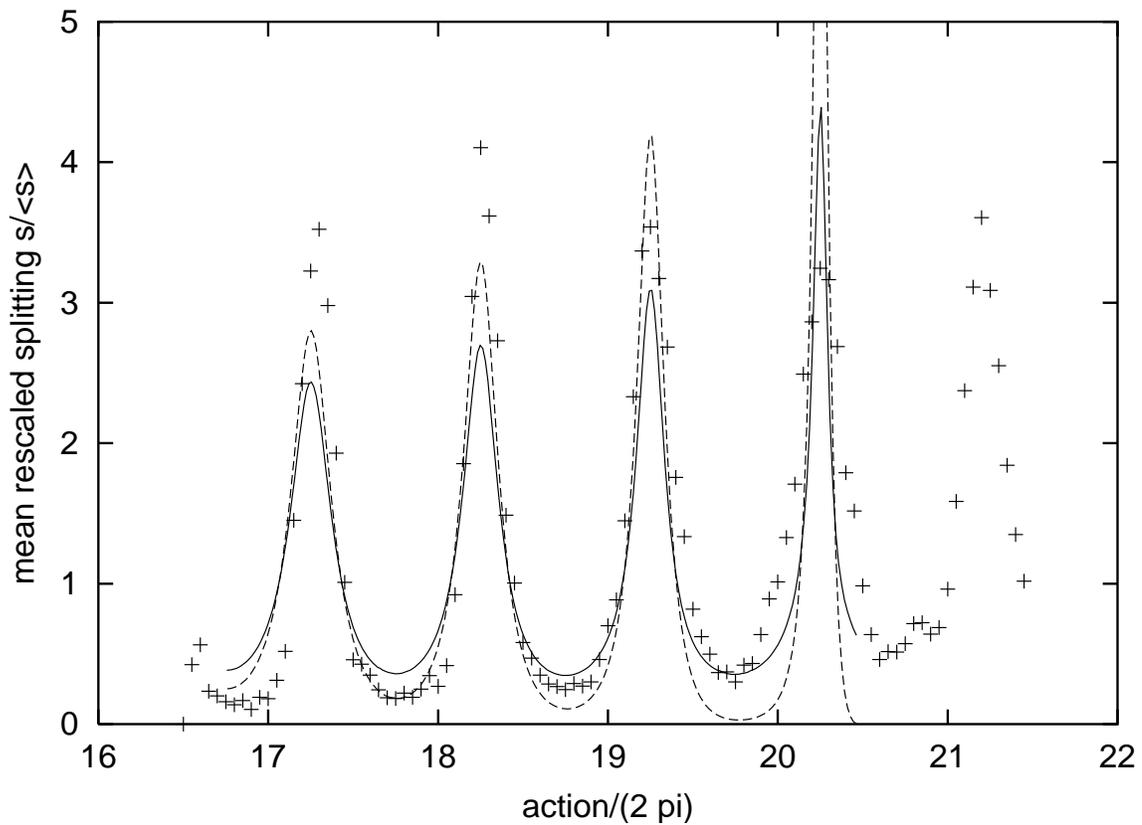,width=6in,bbllx=50pt,bblly=50pt,bburx=540pt,bbury=405pt,clip=}}

\vskip 0.2in

\caption{Mean rescaled splitting {$s/\langle s \rangle$} versus 
action/{$2 \pi$}, data points;
short-time envelope from scar theory using Eq.~(\ref{auto2}), solid line;
short-time envelope from scar theory using Eq.~(\ref{auto1}), dashed line.
}

\label{envelope}
\end{figure}

The connection between scarring and tunneling can be 
tested quantitatively in two ways.
First, scar theory in the semiclassical limit predicts that the 
short-time (smooth) envelope describing the oscillations in the
mean rescaled splitting versus action is given by the
Fourier transform of the autocorrelation function {$A(m)=\langle \phi
|\phi(m)\rangle$}, where {$\phi$} is a Gaussian wave packet (living in the
Poincar\'e section) centered on the horizontal periodic orbit and
{$\phi(m)$} is its iterate after {$m$} bounces. The Gaussian {$\phi$} is
chosen to have the same orientation and aspect ratio
in the {$(y,p_y)$}-plane as the
Gaussian called for by the Creagh--Whelan theory, but linearly
rescaled so as to have area {$h$} as needed for scar theory.
Linearizing the dynamics around
the horizontal periodic orbit we find,
when the Gaussian
wave packet is optimally aligned along the stable and unstable manifolds
of the periodic orbit,
\begin{equation}
A(m)={1 \over {\sqrt{\cosh \lambda m}}} \,,
\label{auto1}
\end{equation}
where {$\lambda$} is the Lyapunov exponent of the periodic orbit.
In the special case of orthogonal stable and unstable manifolds,
a circular Gaussian will be one example of an optimally aligned wave packet.
If the Gaussian is not optimally aligned,
this formula may be generalized as follows:
\begin{equation}
A(m)=2\sqrt{ {{\det(M)} \over {\det(M+(J^{-m})^T M J^{-m})}}} \,,
\label{auto2}
\end{equation}
where {$M$} describes a Gaussian of the form {$({\rm const})\exp(-x^T M x)$}
with {$x=(y,p_y)^T$} representing the coordinates in the surface of section,
and {$J$} is the Jacobian of the Poincar\'e mapping evaluated at the
periodic orbit (${\rm Tr} \; J = 2 \cosh \lambda$).
The matrix {$M$} is computed as specified in the Creagh--Whelan theory
from the monodromy matrix of the complex orbit that begins at the
Poincar\'e section on the right, goes through the barrier, and ends
at the Poincar\'e
section in the left well \cite{CW}.
Here, the Lyapunov exponent {$\lambda$} and Jacobian {$J$} vary with energy
over the range {$-50<E<-9$}. At higher
energies, the trajectory spends less time near the Gaussian at the
center of the well, and thus experiences less deflection, leading to
greater stability, eventually becoming stable for $E>-9$.
The short-time envelope obtained as the Fourier transform 
of {$A(m)$} in either Eq.~(\ref{auto1}) or Eq.~(\ref{auto2}) may be compared 
with the mean rescaled splitting plotted versus action.
As shown in 
Fig.~\ref{envelope}, with either form of the autocorrelation function
we do find peaks in the predicted envelope of splittings at the right
values of action for energies {$E<-9$} (for energies {$E>-9$} the horizontal
periodic orbit becomes stable, so the scar theory does not apply and no 
prediction about the distribution of splittings can be made),
but the heights of the maxima and minima between the peaks are not well 
reproduced. The contrast predicted by Eq.~(\ref{auto2}) is closer to the 
numerical data than that predicted by Eq.~(\ref{auto1}).
The quantitative failure of semiclassical scar theory is
attributable to the
fact that, for our parameter values, the linearizable region around the 
horizontal periodic orbit is not large compared to {$h$}. In fact, the size of
the linearizable region is only about 0.15{$h$} for the energies considered.
Its size, however,
is approximately independent of energy for {$-50 < E < 0$}, and this may
explain the weak dependence of the peak height on energy observed in the
numerical data.
Nevertheless, we see that
not only is scarring associated with larger splittings in the
coarse sense of Fig.~\ref{so} but also the enhancement factor in the
distribution of splittings, as a function of energy or of action,
does oscillate with energy or action
in agreement with the analytical prediction of 
scar theory; only the precise
magnitude of these oscillations remains unexplained
within the present linear theory. 

\begin{figure}

\centerline{
\psfig{file=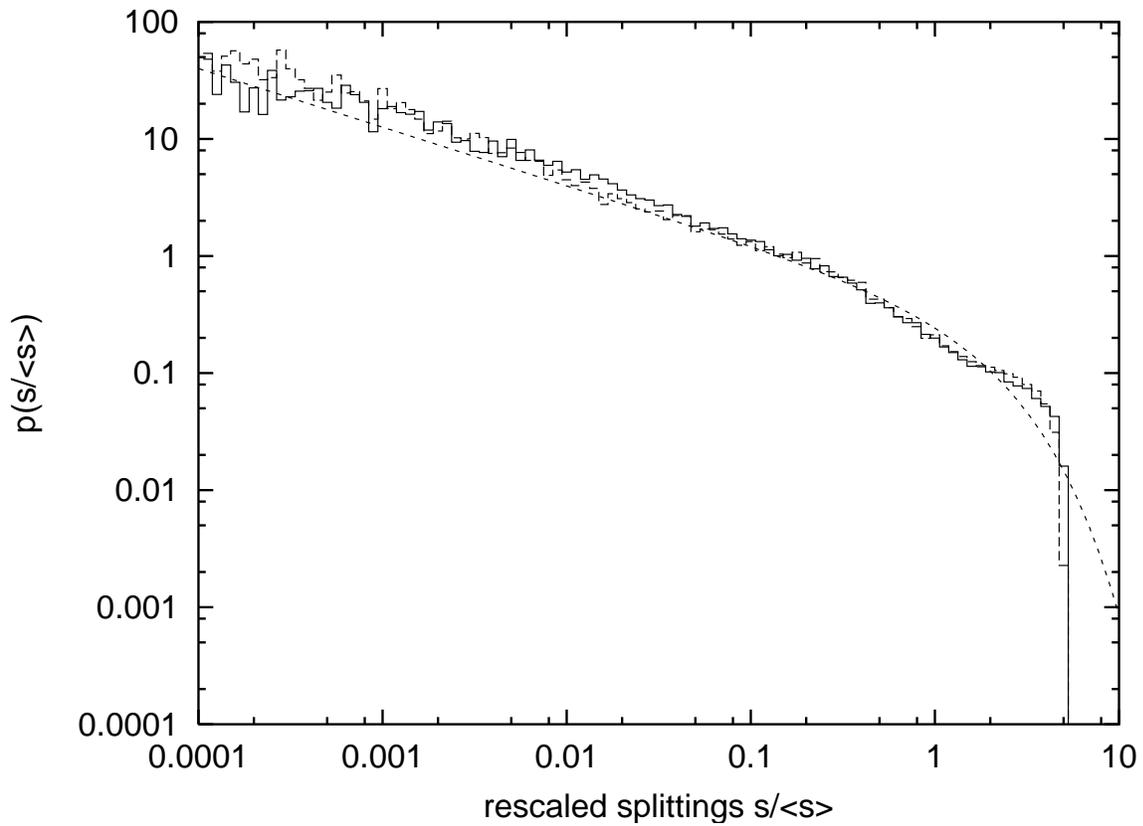,width=6in,bbllx=50pt,bblly=50pt,bburx=540pt,bbury=405pt,clip=}}

\vskip 0.2in

\caption{Distribution of rescaled splittings for numerical data with
{$\sigma=0.5$}, solid histogram; numerical data with {$\sigma=0.63$},
dashed histogram; Porter--Thomas distribution (without correction for
finite {$\hbar$}, see Fig.~\ref{sfhbar}), dashed line.}

\label{sll}
\end{figure}

A second quantitative test
of scar theory in relation to tunneling is to examine
the change in the distribution of splittings upon change in the
Lyapunov exponent. The horizontal periodic orbit can easily be made
more stable by keeping the height of the main Gaussian bump fixed at 150 while
increasing its width. An ensemble of eigenstates and associated splittings
was computed, just as above, for a larger value of the bump width, namely
$0.63$ instead of $0.50$. The expectation
from scar theory would be for the distribution of splittings to have
many more smaller and larger splittings at the resulting smaller Lyapunov 
exponent. At {$\sigma=0.5$}, the horizontal periodic orbit is stable down to
{$E=-9.1$}; the Lyapunov exponent then increases from zero 
with decreasing energy to a value of {$\lambda=2.0$} at {$E=-50$}.
For {$\sigma=0.63$} it is stable down to {$E=-49.8$} and attains only a
value of {$\lambda=0.11$} at {$E=-50$}.
The numerical data, however, 
show no marked difference between the two computations at 
{$\sigma=0.5$} and {$\sigma=0.63$}; see Fig.~\ref{sll}.
The lack of any difference between 
the distributions of splittings despite the difference in stability
is an indication that we are not far enough into the semiclassical
limit (see discussion below). 
In both cases the distribution of rescaled splittings (see the histograms in
Fig.~\ref{sll}) has many more small and large splittings,
and consequently fewer splittings around {$s/\langle s \rangle = 1$},
than a Porter--Thomas
distribution would have (except for the sharp cutoff at 
{$s/\langle s \rangle \ge 5$}, which will be discussed below). 
Thus, the prediction of scar theory that there should be
many more small and large splittings, relative to the prediction of
random matrix theory, is confirmed. Also, the divergence of the
probability distribution near zero splitting in the case of scarring
on the real continuation of the optimal tunneling path differs markedly
from the results, both analytical and numerical, of Creagh and Whelan
\cite{CW2} for the case when the real continuation is not a periodic
orbit, which show a probability distribution tending to zero at zero
splitting. Our numerical results for the scarring case improve on their
statistics and allow us to discern the scar corrections to Porter--Thomas
behavior. In particular, we note the excess of very small splittings;
these correspond to the phenomenon of antiscarring, as seen in 
Fig.~\ref{sa} at actions halfway between values of action given by
the EBK quantization condition for maximal scarring. As studied by
Kaplan \cite{antiscar}, in an open quantum system coupled to the environment
by one channel located on a short unstable periodic orbit, antiscarring
causes the probability to remain in the system at times large compared
to the Heisenberg time to be substantially enhanced relative to the
prediction of random matrix theory.
Therefore, we must expect that antiscarring, which
we have demonstrated now for the case of level splittings in
a smooth chaotic double-well potential, would markedly alter, away from 
random matrix theory predictions, the
distribution of resonance widths in a chaotic metastable
potential, and also the long-time
probability to remain in such a well.

\begin{figure}

\centerline{
\psfig{file=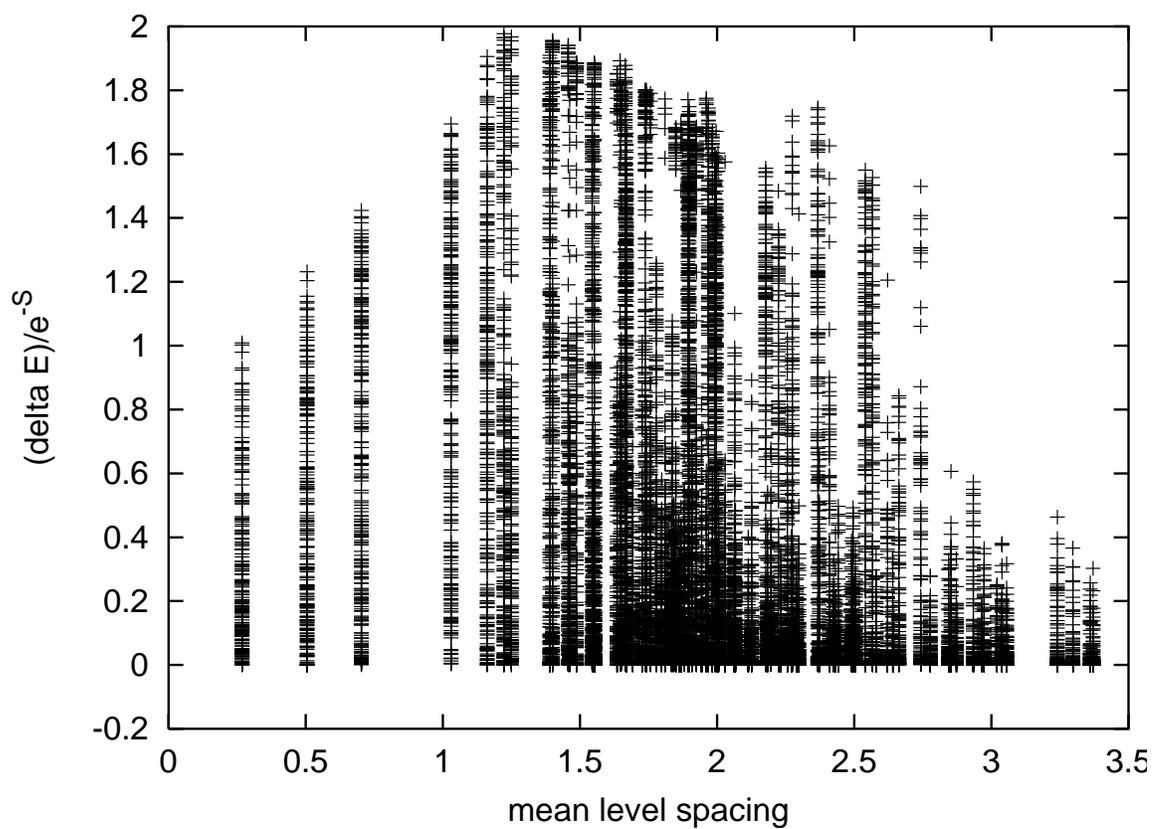,width=6in,bbllx=50pt,bblly=50pt,bburx=540pt,bbury=405pt,clip=}}

\vskip 0.2in

\caption{Plot of {$\Delta E/e^{-S}$} versus mean level spacing {$\Delta$}.}

\label{esvsd}
\end{figure}

In Fig.~\ref{esvsd} we show the relation between the rescaled splitting
{$\Delta E/e^{-S}$}
and the mean level spacing {$\Delta$}, which decreases from about 3
at {$E=-50$} to about 1 at {$E=0$}, aside from some fluctuations.
There is no direct correlation between {$\Delta E/e^{-S}$} and {$\Delta$}, 
thus refuting the intuitive expectation that the tunneling rate should be
proportional only to the rate of attempts to cross the barrier given
by the classical motion, as discussed above in Section \ref{intro}.
In the presence of scarring on the horizontal periodic orbit,
tunneling is enhanced by the tendency to remain near the horizontal 
periodic orbit. 
At energies
for which scarring takes place, the typical wave function
intensity measured using a Gaussian at the
turning point will be enhanced by a factor of $O(1/\lambda)$ compared with
the naive expectation, where
{$\lambda$} is the Lyapunov exponent. At energies for which
antiscarring, a tendency to avoid the horizontal periodic orbit, 
takes place, this typical intensity will be strongly suppressed,
by an amount that is exponentially
small in {$\lambda$} for small {$\lambda$}.
The actual distribution of the rescaled splitting {$\Delta E/e^{-S}$}
versus the level spacing {$\Delta(E)$}
includes the same energy-dependent oscillations
seen in Fig.~\ref{serescaled}, as a function of {$\Delta(E)$} rather than
of {$E$} itself. It is evident, then, that chaotic tunneling in two dimensions
must be thought of as a quantum-coherent phenomenon, in which the probability 
of tunneling through the barrier is greater if one comes back in phase when 
making repeated attempts to cross the barrier, as happens for scarred
eigenfunctions. We also note that the horizontal periodic orbit becomes more
unstable at lower energies, leading to smaller scar peaks in the mean
wave function intensity on the orbit, and thus compensating to some extent 
for an increase in the mean level spacing at lower energies. This may partly
explain the absence a clear trend in the data of Fig.~\ref{esvsd}.

\begin{figure}

\centerline{
\psfig{file=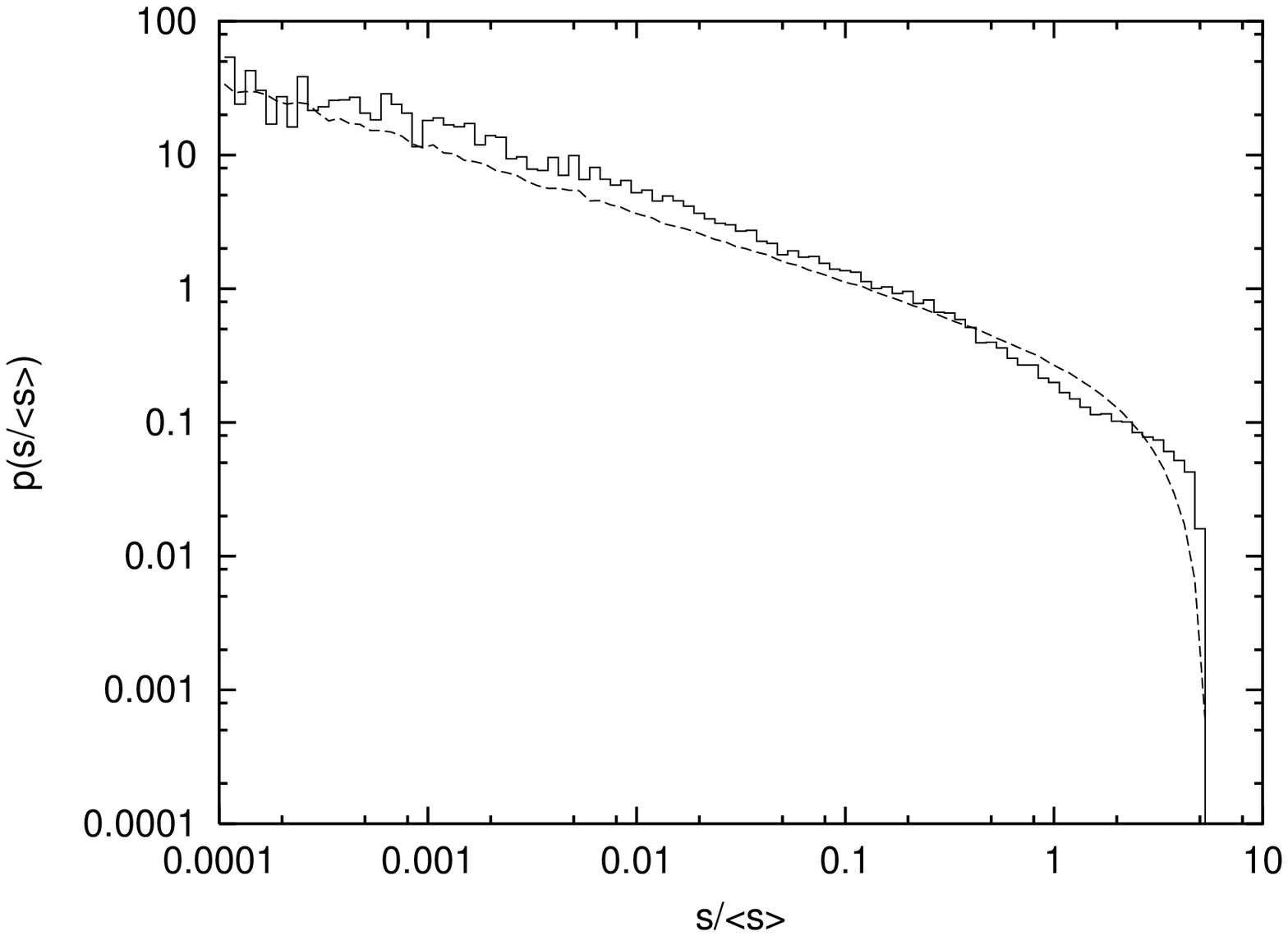,width=6in,bbllx=50pt,bblly=50pt,bburx=540pt,bbury=405pt,clip=}}

\vskip 0.2in

\caption{Distribution of rescaled splittings for the numerical data with
{$\sigma=0.5$}, histogram; Porter--Thomas distribution with finite-{$\hbar$}
correction for {$N=6$}, dashed line.}

\label{sfhbar}
\end{figure}

We now discuss how our data are limited by the fact that we must 
work at finite {$\hbar$}. First, the sharp cutoff at large splittings
in the numerical data relative to the Porter--Thomas distribution
in Fig.~\ref{sll} is a finite-{$\hbar$} effect. This can be understood
as follows. Let the Poincar\'e surface of section have area {$N$} in units
of {$h$}. Then the expected squared overlap {$\langle s \rangle$} of
an eigenstate with a Gaussian test state will be {$1/N$}, because
the test state covers an area {$h$} in phase space while the eigenstate
is, on average,
spread evenly over the entire phase space. Now the cutoff arises from
the fact that no matter how scarred or otherwise localized the eigenstate
is, its overlap with a test state cannot be greater than unity. So
{$s<1$} by construction, or {$s/\langle s \rangle < N$}. Thus the
cutoff increases to infinity in the semiclassical limit (i.e.,
as {$\hbar$} tends to zero), even while {$\langle s \rangle$} itself
is decreasing. Assuming random matrix theory, the modified form of the
Porter--Thomas distribution for finite {$N$} can be computed. One takes
an ensemble of randomly oriented vectors in {$N$} dimensions, normalizes
them so they lie on the unit sphere, and takes {$N$} times the square of the
{$z$}-component. This quantity has mean 1 and a sharp cutoff at {$N$}.
The Porter--Thomas distribution is recovered in the limit {$N \rightarrow
\infty$}. For an analytical form for the Porter--Thomas distribution for
finite {$N$}, see Brody {\it et al.} in Ref.~\cite{BFFMPW},
especially their Eq. (7.10).
In Fig.~\ref{sfhbar} we see that the modified Porter--Thomas
distribution for {$N=6$}, corresponding roughly
to the effective dimension of our Hilbert space,
reproduces the cutoff in the numerical data
of Fig.~\ref{sll}. The scarring corrections (extra splittings at large and
small {$s/\langle s \rangle$} with fewer splittings around 
{$s/\langle s \rangle = 1$}) relative to the modified Porter--Thomas
distribution are still present in Fig.~\ref{sfhbar}.

A second test of the effect of finite {$\hbar$} is to repeat the
calculation at different values of {$\hbar$}. Since we are near the 
computational limit already, we consider only the case of larger
{$\hbar$}. This is done by scaling the coordinates {$(x,y) \rightarrow
(x',y') = (cx,cy)$}, {$0<c<1$}. Under this transformation the potential
becomes
\begin{equation}
V(x',y')={{x'^4} \over {c^4}} - {{x_0^2x'^2} \over {c^2}} +
{{ay'^2} \over {c^2}} + {{\lambda x'^2 y'^2} \over c^4} +
\sum_i b_i e^{-((x'-cx_i)^2+(y'-cy_i)^2)/(c\sigma_i)^2}\,,
\end{equation}
while the kinetic energy remains
\begin{equation}
-{\partial^2 \over {\partial x'^2}}-
{\partial^2 \over {\partial y'^2}}
\end{equation}
since the momenta are not affected by the transformation.
The complete transformation of the Hamiltonian may be regarded as
the product of three transformations: (i) the scaling of coordinates
by a factor of {$c^{1/2}$} and momenta by a factor of {$c^{-1/2}$}, 
which does not change the quantum mechanics, (ii) scaling both coordinates
and momenta by a common factor of {$c^{1/2}$} while also replacing
the Hamiltonian {$H$} by {$cH$}, which preserves the classical mechanics
exactly but is not area-preserving, and thus affects the quantum mechanics
by changing  the effective value of {$\hbar$},
and (iii) scaling the Hamiltonian
by a factor of {$1/c$}, which trivially rescales the spectrum back into the
original range.
The reason we use this transformation is to keep the classical mechanics,
all the
periodic orbits, their stability properties, etc. unchanged as we
change the effective value of
{$\hbar$}, so the results for different values of the effective {$\hbar$}
(which scales as $1/c$)
are directly comparable.

For {$c=0.8$} we find the same oscillations observed previously
in the distribution of
rescaled splittings 
as a function of energy, only now there are four peaks in the range
from {$E=-50$} to {$E=0$} compared to the five that
we saw before, corresponding to
a larger effective value of {$\hbar$} in the new system.
The distribution of splittings is given by the
scarring corrections to the modified Porter--Thomas distribution for
{$N=4$} now, compared to {$N=6$} above.
Thus, the same conclusions continue to hold but with the expected 
modifications for larger {$\hbar$}. This indicates that at {$c=1$}
we are far enough into the semiclassical regime to see characteristic
semiclassical behavior for the locations of the scarring peaks,
if not for their precise heights.

\section{Conclusions}

We have demonstrated that scarring on the real continuation of the
optimal tunneling path, if it is an unstable periodic orbit,
enhances tunneling and thus leads to larger splittings between the
symmetric and antisymmetric in {$x$} eigenfunctions at energies
near the scarring
energies (likewise,
antiscarring in between the scarring energies
leads to smaller splittings). The energy dependence of
the distribution of splittings displays EBK quantization, and the
shape of the smooth envelope is roughly consistent with the prediction of
scar theory, though the magnitude of the oscillations is not quantitatively 
predicted by the simple linearized dynamics; a better understanding of the
shape of the envelope would require extending scar theory to the
non-linear regime. Also, the distribution of splittings is approximately
Porter--Thomas with scarring corrections, as we would expect on the
basis of scar theory combined with the theory of Creagh and Whelan,
discussed in Section \ref{intro}. We do not find, however, the expected 
dependence on the Lyapunov exponent of the horizontal periodic orbit.
This is presumably due to the fact that our calculations do not probe very
far into the semiclassical limit, our well being only a few wavelengths
across in the transverse ($y$) direction. Finite-{$\hbar$} effects cut off the
far tail of the splitting distribution at all energies.

According to Eqs.~(\ref{dewkb}) and (\ref{gwkb}), suitably generalized to
the chaotic double-well potential in two dimensions as discussed in
Section \ref{intro}, the rescaled resonance widths
in a single
metastable well, the potential of which agreed with the double-well
potential we are using for {$x<+x_0/\sqrt{2}$}, would have the same
distribution as the rescaled splittings we have computed. Thus, our results
imply a non-statistical distribution of resonance widths in a
chaotic metastable well. In view of its importance for chemical
physics, this conclusion deserves further investigation.

Finally, we discuss the prospects for many-dimensional systems. If there
exists an unstable periodic orbit near the real continuation of the
optimal tunneling path in a double well
or metastable well, scarring and antiscarring will again play a role.
The only question is whether the degree of instability is small enough
for scarring to be important; for a Lyapunov exponent {$\lambda$} large
compared to unity the short-time envelope approaches the uniform limit of
random matrix theory.
However, as long as the sum of all instability exponents in directions
transverse to the reaction coordinate does not become large, scarring 
effects are expected to appear, just as in the two-dimensional case 
discussed in the present paper.

\section{Acknowledgments}

This research was supported by the National Science Foundation under
Grant No.\ CHE-9610501, by the Department of Energy under Grant
No.\ DE-FG03-00-ER41132, and by the Institute for Theoretical Atomic
and Molecular Physics at the Harvard-Smithsonian Observatory.

\newpage

\newpage

\newpage

\newpage

\newpage

\newpage

\newpage

\newpage

\newpage

\newpage

\newpage

\begin{references}

\bibitem[*]{bies-email}%
    Electronic address: {\tt bies@fas.harvard.edu}.

\bibitem{chemical}
R. Hernandez, W.~H. Miller, C. Bradley Moore and W.~F. Polik,
{\it J. Chem. Phys.} {\bf 99}, 950 (1993);
W.~H. Miller, R. Hernandez, C.~B. Moore and W.~F. Polik,
{\it J. Chem. Phys.} {\bf 93}, 5657 (1990);
P.~J. Robinson and K.~A. Holbrook, {\it Unimolecular Reactions} 
(Wiley-Interscience, New York, 1972);
J.~I. Steinfeld, J.~S. Francisco and W.~L. Hase, {\it Chemical Kinetics
and Dynamics} (Prentice Hall, Englewood Cliffs, N.J., 1989). 

\bibitem{nuclear}
J.~E. Lynn, {\it The Theory of Neutron Resonance Reactions} (Oxford, 1968).

\bibitem{PT}
C.~E. Porter and R.~G. Thomas, {\it Phys. Rev.} {\bf 104}, 483 (1956).

\bibitem{mesoscopic}
L.~P. Kouwenhoven, C.~M. Marcus, P.~L. Mceuen, S. Tarucha, R.~M.
Westervelt and N.~S. Wingreen, `Electron transport in quantum dots,'
in {\it Mesoscopic Electron Transport}, ed. L.~L. Sohn, L.~P. 
Kouwenhoven and G. Sch\"on, (Kluwer, 1997), pp. 105--214;
C.~M. Marcus, R.~M. Westervelt, P.~F. Hopkins and A.~C. Gossard,
{\it Chaos} {\bf 3}, 643 (1993).

\bibitem{tunneling-diode}
T.~M. Fromhold, L. Eaves, F.~W. Sheard, M.~L. Leadbeater,
T.~J. Foster and P.~C. Main, {\it Phys. Rev. Lett.} {\bf 72}, 2608 (1994);
P.~B. Wilkinson, T.~M. Fromhold, L. Eaves,  F.~W. Sheard,
N. Miura and T. Takamasu, {\it Nature} {\bf 380}, 608 (1996);
T.~M. Fromhold, P.~B. Wilkinson, F.~W. Sheard, L. Eaves,
J. Miao and G. Edwards, {\it Phys. Rev. Lett.} {\bf 75}, 1142 (1995).

\bibitem{wkb}
D. Bohm, {\it Quantum Theory} (Prentice-Hall, New York, 1951), pp. 264--295. 

\bibitem{CW}
S.~C. Creagh and N.~D. Whelan, {\it Ann. Phys.} {\bf 272}, 196 (1999).

\bibitem{CW2}
S.~C. Creagh and N.~D. Whelan, {\it Phys. Rev. Lett.} {\bf 84}, 4084 (2000).

\bibitem{dvr}
D.~O. Harris, G.~G. Engerholm and W.~D. Gwinn, {\it J. Chem. Phys.} {\bf 43},
1515 (1965); for more recent work see J.~V. Lill, G.~A. Parker and J.~C.
Light, {\it Chem. Phys. Lett.} {\bf 89}, 483 (1982); J.~C. Light,
I.~P. Hamilton and J.~V. Lill, {\it J. Chem. Phys.} {\bf 82}, 1400 (1985).

\bibitem{KH}
L. Kaplan and E.~J. Heller, {\it Ann. Phys.} {\bf 264}, 171 (1998);
see also the review: L. Kaplan, {\it Nonlinearity} {\bf 12}, R1 (1999)
and references therein.

\bibitem{AF}
O. Agam and S. Fishman, {\it Phys. Rev. Lett.} {\bf 73}, 806 (1994);
O. Agam and S. Fishman, {\it J. Phys. A} {\bf 26}, 2113 (1993).

\bibitem{LK}
L. Kaplan, {\it Phys. Rev. Lett.} {\bf 80}, 2582 (1998).

\bibitem{SCdVZ}
A.~I. Schnirelman, {\it Usp. Mat. Nauk.} {\bf 29}, 181 (1974);
Y. Colin de Verdiere, {\it Commun. Math. Phys.} {\bf 102}, 497 (1985);
S. Zelditch, {\it Duke Math. J.} {\bf 55}, 919 (1987);
S. Zelditch and M. Zworski, {\it Commun. Math. Phys.} {\bf 175}, 673 (1996).

\bibitem{antiscar}
L. Kaplan, {\it Phys. Rev. E} {\bf 59}, 5325 (1999).

\bibitem{BFFMPW}
T.~A. Brody, J.~Flores, J.~B. French, P.~A. Mello, A. Pandey and S.~S.~M. Wong,
{\it Rev. Mod. Phys.} {\bf 53}, 385 (1981).

\end{references}
\end{document}